\documentclass[aps,floatfix,prd,twocolumn,nofootinbib]{revtex4}
\usepackage{graphicx}
\usepackage{dcolumn}
\usepackage{bm}
\usepackage{amsmath}
\usepackage{float}
\usepackage{lipsum}
\usepackage{color}
\usepackage{times}
\usepackage{textcomp}
\usepackage{latexsym}
\usepackage{braket}
\usepackage[hidelinks]{hyperref}
\usepackage{acro}

\hypersetup{colorlinks = true,  allcolors = blue}

\DeclareAcronym{gw}{
  short = GW ,
  long = gravitational wave ,
  short-plural = ,
}
\DeclareAcronym{bh}{
  short = BH ,
  long = black hole ,
  short-plural = ,
}
\DeclareAcronym{psd}{
  short = PSD ,
  long = power spectral density ,
  short-plural = ,
}
\DeclareAcronym{snr}{
  short = SNR ,
  long = signal-to-noise ratio,
  short-plural = ,
}
\DeclareAcronym{ligo}{
  short = LIGO ,
  long = Laser Interferometer Gravitational-Wave Observatory ,
  short-plural = ,
}
\DeclareAcronym{eh}{
  short = EH ,
  long = The event horizon ,
  short-plural = ,
}
\DeclareAcronym{ah}{
  short = AH ,
  long = the apparent horizon ,
  short-plural = ,
}
\DeclareAcronym{pn}{
  short = PN ,
  long = Post-Newtonian,
  short-plural = ,
}
\DeclareAcronym{ppn}{
  short = PPN ,
  long = Parameterized Post-Newtonian,
  short-plural = ,
}

\begin{document}

\title{Constraining black-hole horizon effects by LIGO-Virgo detections of inspiralling binary black holes}
\author{Kwun-Hang Lai}
\author{Tjonnie Guang Feng Li}
\affiliation{
Department of Physics,
The Chinese University of Hong Kong,
Sha Tin,
Hong Kong.}

\date{\today}

\begin{abstract}
\noindent 
General relativity predicts mass and spin growth of an inspiralling black hole due to an energy-momentum flux flowing through the black-hole horizon. The leading-order terms of this horizon flux introduce 2.5 and 3.5 post-Newtonian corrections to inspiral motions of binary black holes. The corrections may be measurable by gravitational waves detectors. Since the proper improvements to general relativity is still a mystery, it is possible that the true modified gravity theory introduces negligible direct corrections to the geodesics of test masses, while near horizon corrections are observable. We introduce a parameterization to describe arbitrary mass and spin growth of inspiralling black holes. Comparing signals of gravitational waves and a waveform model with parameterized horizon flux corrections, deviations from general relativity can be constrained. We simulate a set of gravitational waves signals following an astrophysical distribution with horizon flux modifications. Then, we perform a Bayesian analysis to obtain the expected constraints from the simulated response of the Advanced LIGO-Virgo detector network to the simulated signals. We show that the constraint can be improved by stacking multiple detections. The constraints of modified horizon flux can be used to test a specific class of modified gravity theories which predict dominant corrections near black-hole horizons over other types of corrections to general relativity. To support Hawking's area theorem at 90\% confidence level, over 10000 LIGO-Virgo detections are required. Within the lifetime of LIGO and Einstein Telescope, a future ground-based gravitational wave detector, near horizon corrections of modified gravity theories are potentially detectable if one of the modified gravity theory is true and the theory predicts a strong correction.
\end{abstract}

\maketitle

\section{Introduction}
\noindent 
In the observable universe, regions near black-hole horizons contain the most extreme spacetimes. The strong curvature nature of black-hole horizons provides a stage for interesting physical phenomena, both classically and quantum mechanically. A set of thermodynamics laws governing the evolution of black hole horizons is predicted by general relativity and quantum physics~\citep{jacobsonintroductory, mensky1998relativistic}. It is natural to ask if we can observe deviations from general relativity in such a strong curvature regime. In fact, modified gravity theories can predict significantly different features of black holes or black-hole-like objects. For example, a specific type of scalar-tensor gravity predicts black holes with zero temperature~\citep{bronnikov2016black}. It is possible to construct theories which predict observable deviations from general relativity in the near horizon regime, while effects on the orbital motion of test particles are negligible. Examples include scalar-tensor-vector gravity thermodynamics and quantum corrections to black-hole entropy. Those corrections may lead to violation of the area theorem. Constraining near horizon modifications with observational data can rule out some of the theories in this category and hint at the correct form of gravity.

Currently, X-ray binary systems~\citep{mcclintock2003black,remillard2006x}, measurements of stellar orbital motions around supermassive black holes~\citep{ghez2005stellar,ghez2008measuring,kormendy2013coevolution} and gravitational waves~\citep{abbott2016observation,abbott2016gw151226,scientific2017gw170104,abbott2017gw170814,abbott2016binary} are the main ways to observe black holes, and provide the possibility of observing the physics of  black-hole horizons. X-ray binary systems require complicated models to describe atomic processes~\citep{schatz2006x}, which makes it hard to extract the information of horizons. The Event Horizon Telescope is a telescope network that observes astronomical objects near supermassive black hole horizons directly~\citep{castelvecchi2017hunt}. In contrast, we propose an alternative method to extract the physics of horizons through observing binary black-hole merger by gravitational waves.

A gravitational wave emitted by a binary black-hole merger consists of three phases: inspiral, merger, ringdown. The properties of black hole horizons during the merger, including the area theorem, are discussed in~\citep{gupta2018dynamics}. To extract the information of black hole horizons from a gravitational wave detection, Cabero et al.~\citep{cabero2017observational} proposed an observational test by checking the consistency between the initial black-hole area in the inspiral phase and the final black-hole area in the ringdown phase. In this paper, we propose another way to extract the evolution of black-hole horizons from the inspiral phase only. 

In general relativity, the inspiral motion of a binary black holes can be described by \ac{pn} formalism~\citep{blanchet2014gravitational}. The inspiral motion and the corresponding gravitation waves in modified gravity theories can be described under parameterized frameworks including the \ac{ppn}~\citep{will2014confrontation} and the Parameterized Post-Einsteinian~\citep{scientific2016tests} formalism, where deviations from general relativity are parameterized by a set of independent variables. Through comparing parameterized waveforms and gravitational waves detected by \ac{ligo}, parameters of the parameterized waveforms can be constrained by Bayesian analysis~\citep{li2012towards,cornish2011gravitational}. There are no significant deviations from general relativity found so far~\citep{scientific2016tests,abbott2016binary}. 

Waveforms in the \ac{pn} formalism can be derived without considering the internal structure of the objects, which is known as the effacement principle~\cite{maggiore2008gravitational}. For the purpose of detecting gravitational waves, the effacement principle is sufficient~\cite{boyle2009comparison}. However, modified gravity theories may predict observable corrections near black-hole horizons, while direct effects on geodesics of black holes are negligible. Different phenomena have been investigated to understand the effects of near-horizon corrections~\citep{barack2018black}, including the tidal Love number~\citep{cardoso2017testing,berti2018extreme}, echoes~\citep{cardoso2017observational,cardoso2017tests}, and tidal heating with LISA binaries~\citep{maselli2018probing}. In this paper, we investigate the constraint on a modified horizon flux\footnote{Horizon flux is another name for the tidal heating effect.}, predicted by some modified gravity theories, by simulating responses of an Advanced LIGO-Virgo detector network to waveforms in a modified \ac{pn} formalism. The constraint can be used to test those modified gravity theories with dominating corrections near black-hole horizons. The methodology has a similar philosophy to~\citep{maselli2018probing}, but we focus on the stellar-mass black-hole binaries which are detectable by the Advanced LIGO-Virgo detector network other than the supermassive black-hole considered in~\citep{maselli2018probing}. Since the surface gravity on a spherically-symmetric stellar-mass black hole is stronger than that on a spherically-symmetric supermassive black hole, some modified gravity theories may predict stronger near-horizon corrections for a stellar-mass black hole, and thus the corrections may be better constrained by the Advanced LIGO-Virgo detector network.

This paper is structured as follow: Sec.~\ref{sec:modsection} describes the origin of the horizon flux and the parameterized modification of the flux; Sec.~\ref{sec:bayesian} shows the simulated constraints of the parameterized modification; and Sec.~\ref{sec:dis} discusses the limitations of the methodology and points to future research.


\section{Parametrized horizon effects in gravitational waveforms} \label{sec:modsection}
\noindent 
In this section, we review the horizon flux into a black hole, derived by Alvi~\citep{alvi2001energy}, Poisson~\citep{poisson2004absorption}, and Chatziioannou et al.~\citep{chatziioannou2016improved,chatziioannou2013tidal}. Assuming that the spacetime around a black-hole horizon deviates from general relativity, we introduce a parameterization to model the effect of the horizon flux to a gravitational waveform. The applications of this parameterization are also discussed.

\subsection{Configuration and notations}
\noindent
We focus on the inspiral phase of a binary black hole coalescence, which allows us to separate and study the black-hole horizon effects. Black holes $i$ ($i = 1,2$) are described by mass $m_{i}$ and angular momentum $J_{i}$, where $m_{1} < m_{2}$, and where we require $J_{i}$ to stay parallel/anti-parallel to the orbital angular momentum for simplicity, which is sufficient to cover most of the sources to which LIGO is sensitive~\citep{harry2016searching}. 
The dimensionless spin parameters are defined as $\chi_{i} = J_{i}/m_{i}^{2}$. For convenience, we show some of the equations in total mass $M = m_{1} + m_{2}$ and symmetric mass ratio $\nu = m_{1}m_{2}/M^{2}$. The \ac{pn} velocity is denoted by $x = (\pi M f)^{1/3}$, where $f$ is the frequency of the orbital motion. Throughout the paper, we use the geometrized units $c = G = 1$. 

Each black hole is bounded by a few horizons. \ac{eh} and \ac{ah} are two of the most important horizons of black holes. \ac{eh} is defined as the boundary of a region where no null curve can reach future null infinity~\citep{booth2005black}, while \ac{ah} is defined as the outermost marginally trapped surface~\citep{schnetter2006introduction}. During the inspiral phase of a binary black hole coalescence, we show that \ac{eh} and \ac{ah} are indistinguishable by the \ac{ligo} detectors in Appendix~\ref{sec:aheh}. As a result, we do not discriminate between \ac{ah} and \ac{eh} cross-section areas in a time slice, the horizon areas of the two black holes are denoted as $A_{i}$.

\subsection{Growth of black hole area, mass and spin}
\noindent
By adopting the ingoing Kerr coordinates $(v,r,\theta,\phi)$, the growth of black hole horizon can be written as~\citep{poisson2004absorption}
\begin{equation}
\frac{dA_{i}}{dv} = \oint \Theta dS,
\end{equation}
where $\Theta$ denotes the expansion scalar which quantifies how the black-hole area changes, and $dS$ denotes the differential surface area of the horizon.  

The first law of black hole thermodynamics and the mode decomposition of matter fields can be written as (see~\citep{poisson2004absorption} for details)

\begin{equation} \label{eq:firstlaw}
\frac{\kappa}{8\pi} \langle \dot{A_{i}} \rangle = \langle \dot{m_{i}} \rangle - \Omega_{H} \langle \dot{J_{i}} \rangle,
\end{equation} 
\begin{equation}
\langle \dot{m_{i}} \rangle _{\mu,\omega} = \frac{\omega}{\mu} \langle \dot{J_{i}} \rangle _{\mu,\omega},
\end{equation}
where  $\Omega_{H}$ is the angular velocity of the unperturbed black hole, the brackets denote accumulative growths while eliminating the highly oscillating modes, and note that $\mu$ here refer to an index of the Fourier modes. After solving the Teukolsky's equation, we have~\citep{chatziioannou2016improved}

\begin{equation} \label{eq:spin}
\left \langle \frac{dJ_{i}}{dx} \right \rangle = (\Omega_{H} - \Omega)C'_{x},
\end{equation} 

\begin{equation} \label{eq:mass}
\left \langle \frac{dm_{i}}{dx} \right \rangle = \Omega (\Omega_{H} - \Omega)C'_{x},
\end{equation}

\begin{equation} \label{eq:area}
\left \langle \frac{dA_{i}}{dx} \right \rangle = - \left( \frac{8\pi}{\kappa} \right) (\Omega_{H} - \Omega)^{2} C'_{x},
\end{equation}
where $\Omega$ is angular velocity of the tidal field created by its companion black hole respectively. $C'_{x}<0$~\citep{chatziioannou2016improved} is a function of $m_{1}$, $m_{2}$, $\chi_{1}$, $\chi_{2}$ and $x$. Together with $dx/dt > 0$~\citep{buonanno2009comparison}, Eq.~\eqref{eq:area} implies Hawking's area theorem $\langle dA_{i}/dt \rangle > 0$.  

Eqs.~\eqref{eq:spin} and~\eqref{eq:mass} can be interpreted as the angular momentum flux and mass flux into a Kerr black-hole horizon. Mass flux and angular momentum flux flow into a general horizon are well defined, though it is derived in a different approach~\citep{ashtekar2004isolated}. 

\subsection{2.5PN and 3.5PN horizon terms in frequency domain} 
\noindent
Eqs.~\eqref{eq:spin} and~\eqref{eq:mass} can be integrated in the \ac{pn} barycentric frame to give a 2.5\ac{pn} and 3.5\ac{pn} phase term (Appendix~\ref{modifiedflux}). We follow the logic of~\citep{isoyama2017post} to derive the effect of the horizon flux in the TaylorF2 formalism\footnote{While we agree on the 2.5\ac{pn} phase term, we compute a slightly different 3.5\ac{pn} phase term comparing to Eq. (4.40) in~\citep{isoyama2017post}. However, the difference is negligible for the cases considered in this paper, see the discussion in Appendix~\ref{modifiedflux}.}. 

In the frequency domain, the TaylorF2 gravitational waveform $h(f)$ with the horizon effect can be written as 

\begin{equation}
h(f) = A^{F2}(f) e^{-i [\Psi^{F2}(f) + \Psi_{H}^{F2}(f)]},
\end{equation}
where $A^{F2}(f)$ and $\Psi^{F2}(f)$ are the amplitude and the phase of the TaylorF2 waveform respectively without the horizon effect (we refer to~\citep{buonanno2009comparison,mishra2016ready} for the 3.5\ac{pn} aligned spins TaylorF2 waveform without the horizon terms), and $\Psi_{H}^{F2}(f)$ corresponds to the horizon effect phase term. 

The horizon flux introduces extra 2.5\ac{pn} ($\Psi^{F2}_{H,5}$) and 3.5\ac{pn} ($\Psi^{F2}_{H,7}$) phase terms. The phase terms are given by 

\begin{equation} \label{eq:F2}
\begin{aligned}
\Psi^{F2}_{H} ={}& \left[1 + 3 \ln \left( \frac{x}{x_{\text{reg}}} \right) \right] \Psi^{F2}_{H,5} + x^{2} \Psi^{F2}_{H,7},
\end{aligned}
\end{equation}
where
\begin{equation} \label{eq:F2term}
\begin{aligned}
\Psi^{F2}_{H,5} ={}&  C_{5\alpha 1} + C_{5\alpha 2},\\
\Psi^{F2}_{H,7} ={}& C_{7\alpha 1} + C_{7\alpha 2} + C_{7\beta 1} + C_{7\beta 2},
\end{aligned}
\end{equation}
and where we ignore the effect of spins on the innermost stable orbit and choose the \ac{pn} velocity of the Schwarzschild innermost stable orbit $x_{\text{ISCO}} = x_{\text{reg}} =  1/\sqrt{6}$ as the stopping condition of the waveform. $C_{5\alpha 1}, C_{5\alpha 2}, C_{7\alpha 1}, C_{7\alpha 2}, C_{7\beta 1}, C_{7\beta 2}$ are given in Appendix~\ref{modifiedflux}.

\subsection{Modified mass and spin flux} \label{ssec:modpara}
\noindent
The variation of mass and spin produces observable effects in the inspiral phase. To test any theory which deviates from general relativity in the near horizon regime, we insert mass-growth parameters $\alpha_{1}, \alpha_{2}$ and spin-growth parameters $\beta_{1}, \beta_{2}$ to parameterize the mass and spin flux deviations on black hole 1,2 respectively:
\begin{equation}
\left \langle \frac{dm_{1}}{dx} \right \rangle  \rightarrow (1+\alpha_{1}) \left \langle \frac{dm_{1}}{dx} \right \rangle,
\end{equation}

\begin{equation}
\left \langle \frac{dm_{2}}{dx} \right \rangle  \rightarrow (1+\alpha_{2}) \left \langle \frac{dm_{2}}{dx} \right \rangle,
\end{equation}

\begin{equation}
\left \langle \frac{dJ_{1}}{dx} \right \rangle  \rightarrow (1+\beta_{1}) \left \langle \frac{dJ_{1}}{dx} \right \rangle,
\end{equation}

\begin{equation}
\left \langle \frac{dJ_{2}}{dx} \right \rangle  \rightarrow (1+\beta_{2}) \left \langle \frac{dJ_{2}}{dx} \right \rangle,
\end{equation} 
where the unmodified flux follows Eqs.~\eqref{eq:spin} and~\eqref{eq:mass}. In general, $\alpha_{1}, \alpha_{2}, \beta_{1}, \beta_{2}$ depend on $x$. In the \ac{pn} framework, $\alpha_{1}, \alpha_{2}, \beta_{1}, \beta_{2}$ can be expressed as polynomial series of $x$. To investigate the simplest case of the horizon flux modification, we restrict $\alpha_{1}, \alpha_{2}, \beta_{1}, \beta_{2}$ to be constant throughout the paper.

With the parameterized mass and spin flux modifications, the modified TaylorF2 2.5\ac{pn} and 3.5\ac{pn} horizon phase terms are given by

\begin{equation} \label{eq:modF2}
\begin{aligned}
\Psi^{F2}_{H} ={}& \left[1 + 3 \ln \left( \frac{x}{x_{\text{reg}}} \right) \right] \Psi^{F2}_{H,5}(\alpha_{1},\alpha_{2}) \\
&+ x^{2} \Psi^{F2}_{H,7} (\alpha_{1},\alpha_{2},\beta_{1},\beta_{2}),
\end{aligned}
\end{equation}

\begin{equation} \label{eq:57com}
\begin{aligned}
\Psi^{F2}_{H,5} ={}& (1+\alpha_{1}) C_{5\alpha 1} + (1+\alpha_{2}) C_{5\alpha 2}, \\
\Psi^{F2}_{H,7} ={}& (1+\alpha_{1}) C_{7\alpha 1} + (1+\alpha_{2}) C_{7\alpha 2} \\
&+ (1+\beta_{1}) C_{7\beta 1} + (1+\beta_{2}) C_{7\beta 2},
\end{aligned}
\end{equation}
where $C_{5\alpha 1}, C_{5\alpha 2}, C_{7\alpha 1}, C_{7\alpha 2}, C_{7\beta 1}, C_{7\beta 2}$ are given in~\eqref{eq:fluxcoe}.

$\Psi^{F2}_{H,5}$ is a function of $\alpha_{1}$ and $\alpha_{2}$, while $\Psi^{F2}_{H,7}$ is a function of $\alpha_{1}$, $\alpha_{2}$, $\beta_{1}$ and $\beta_{2}$. When $\alpha_{1} = \alpha_{2} =  \beta_{1} = \beta_{2} = 0$, Eq.~\eqref{eq:modF2} reduces to the horizon terms in general relativity, given by Eqs.~\eqref{eq:F2} and~\eqref{eq:F2term}.

The TaylorF2 waveform with the parameterized phase terms is utilized to analyze the constraint on the horizon effect in Sec.~\ref{sec:bayesian}. Even though the TaylorF2 model can accurately describe the early inspiral phase~\citep{buonanno2009comparison}, it fails to capture the higher frequency late-inspiral-merger-ringdown phase~\citep{boyle2009comparison}. For investigative purposes, we use TaylorF2 for both signal simulations and data analysis instead of analyzing real signals (such as GW150914~\citep{abbott2016observation}) to avoid misinterpretation of the errors caused by the inaccurate late-inspiral-merger-ringdown band in the TaylorF2 model as a violation of general relativity. However, to test general relativity with real \ac{ligo} signals, a more detailed study on the potential systematic error induced by the inaccuracy of the model is required. With the simplifying assumption in this paper, the results can be interpreted as an optimistic estimation compared to an actual study on real signals.

\subsection{Application of the horizon effect parameterization} \label{ssec:application}
\noindent
If a theory predicts observable deviations from general relativity around black-hole horizons, these deviations can be parameterized by non-zero $\alpha_{1}, \alpha_{2}, \beta_{1}, \beta_{2}$. Furthermore, if the theory predicts negligible direct effects on geodesics of test masses, $\alpha_{1}, \alpha_{2}, \beta_{1}, \beta_{2}$ are the leading order measurable deviations. 

To quantify the requirement where corrections from horizon effects dominate, we consider a point-mass binary system in a modified gravity theory. The orbital energy $E_{\text{orb}}$ can be expressed as a deviation $\delta$ from the \ac{pn} energy $E_{\text{PN}}$,

\begin{equation}  
E_{\text{orb}} = E_{\text{PN}} (1 + \delta).
\end{equation}
The dominating order of $E_{\text{PN}}$ is proportional to the orbital separation $1/r$, where $r$ is proportional to $1/x^{2}$. If $\delta$ can be expressed in a power law, i.e., $\delta \propto 1/r^{n} + O(1/r^{n+1})$, where $n>3.5$, then $\delta$ has no effect on the 3.5\ac{pn} waveform. Therefore, the lowest order correction from the horizon effect can become the leading order correction to general relativity. 

To relate the parameterization to a specific modified gravity theory, the horizon effect should be derived from the theory. Following the philosophy of deriving the horizon effect in general relativity~\citep{poisson2004absorption}, the metric perturbation equation, the gravitational flux flowing into a black hole horizon, and the black hole thermodynamics laws are potentially essential to derive the horizon effect in a modified gravity theory. However, horizons may not exist in some modified gravity theories~\citep{shankar2017horizonless}, which means the derivation process of the horizon effect may not be valid in those theories. 

If we adopt the parameterization $\alpha_{1} = \alpha_{2} =  \beta_{1} = \beta_{2} = \alpha$, then deviations from general relativity including violation of black-hole area theorem, scalar-tensor-vector gravity thermodynamics and quantum corrections to black-hole entropy can be expressed in terms of $\alpha$ (see Table~\ref{tab:theory} and Appendix~\ref{sec:paraexamples}). Since the complete modification due to a modified gravity theory are not included, Table~\ref{tab:theory} suggests some signatures of the modification instead of providing a precise description of the waveform in a modified gravity theory.

\begin{table}[]
\caption{Expressions of the area theorem (Appendix~\ref{ssec:areatheorem}), scalar-tensor-vector gravity thermodynamics (Appendix~\ref{ssec:STVG}) and quantum gravity correction to black-hole entropy (Appendix~\ref{ssec:QG}) in terms of $\alpha$. Symbols $\alpha'$, $\Xi$ are the independent parameters in the corresponding models, while $A$ is the area of a black hole.}
\centering
\begin{tabular*}{\linewidth}{@{\extracolsep{\fill}}l l l}
\hline \hline \noalign{\vskip 1mm} 
Theory                                             & $\alpha$                                          \\ \hline \\
Area theorem~\citep{chrusiel2001regularity}        & $\geq -1$                                         \\ \\
Scalar-Tensor-Vector Gravity thermodynamics~\citep{mureika2016black}        & $= - \frac{\gamma}{1+\gamma+\sqrt{1+\gamma}}$  \\ \\
Quantum corrections to black-hole entropy~\citep{el2016quantum}              & $= \frac{256\pi^{2} \Xi}{A}$                      \\ \\ \hline \\     
\hline \hline
\end{tabular*}
\label{tab:theory}
\end{table}

\section{Constraints on horizon effects} \label{sec:bayesian}
\noindent
To study the constraints on $\alpha_{1}, \alpha_{2}, \beta_{1}, \beta_{2}$, we simulate waveforms and analyze the simulated data using the nested sampling~\citep{skilling2004nested} algorithm in \texttt{LALInference}~\citep{veitch2015parameter}. Motivated by Sec.~\ref{ssec:application}, we focus on the minimal parameterization $\alpha_{1} = \alpha_{2} =  \beta_{1} = \beta_{2} = \alpha$. The following binary black hole parameters are free parameters in the inference process: $\alpha$, masses, aligned/anti-aligned (with orbital angular momentum) spins, inclination, angular sky location and distance. Noise is simulated assuming that all three Advanced LIGO-Virgo detectors are operational at their design sensitivity~\citep{aasi2015advanced,TheVirgo:2014hva}. To estimate the constraints on horizon effects from multiple LIGO-Virgo detections, multiple $\alpha = 0$ (without modification of general relativity) events are simulated and analyzed. 

\subsection{Dependence between constraints of $\alpha$ and binary black holes parameters} \label{ssec:dependency}
\begin{figure}
 \includegraphics[width=0.975\columnwidth]{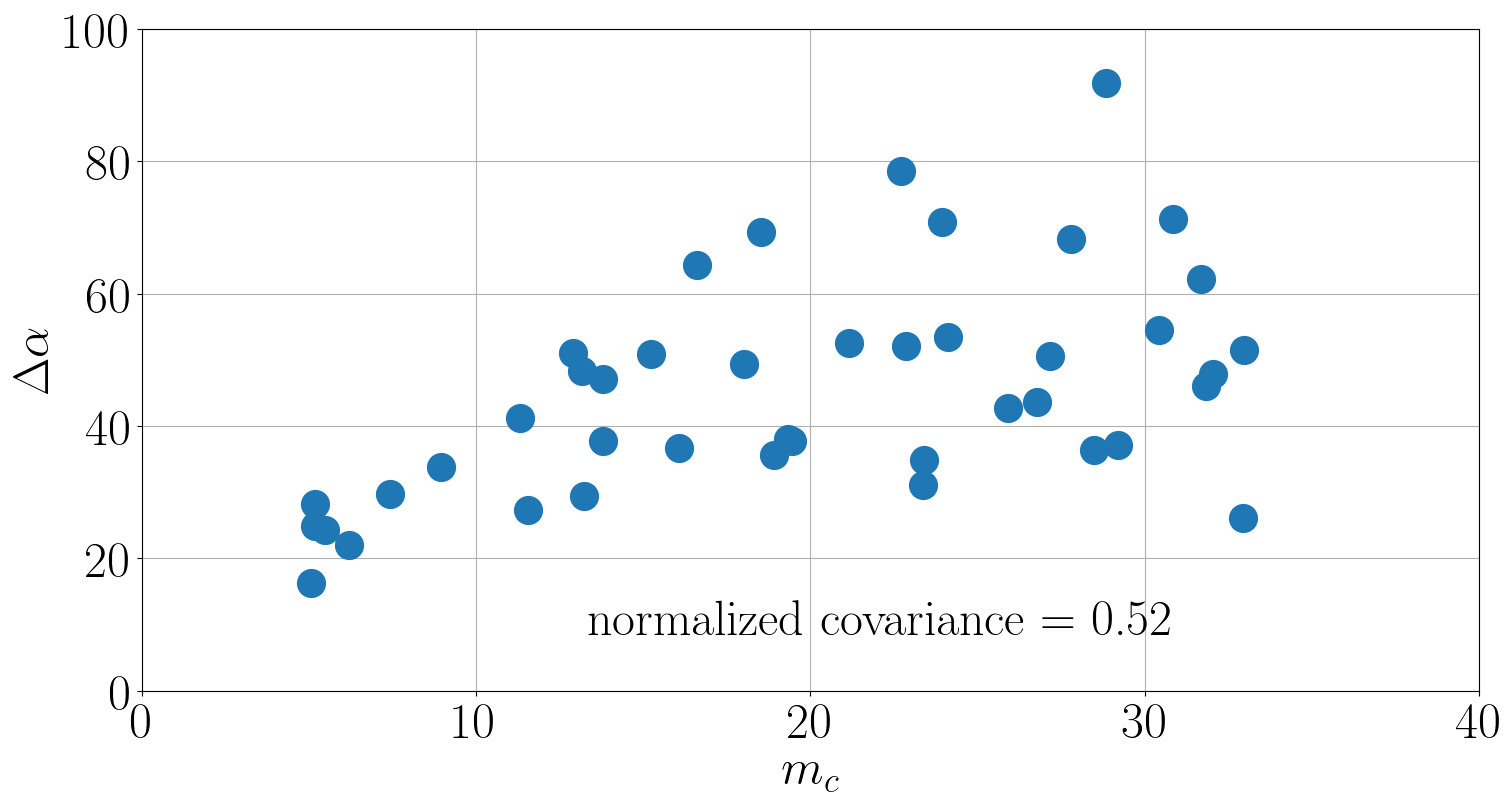}
 \includegraphics[width=\columnwidth]{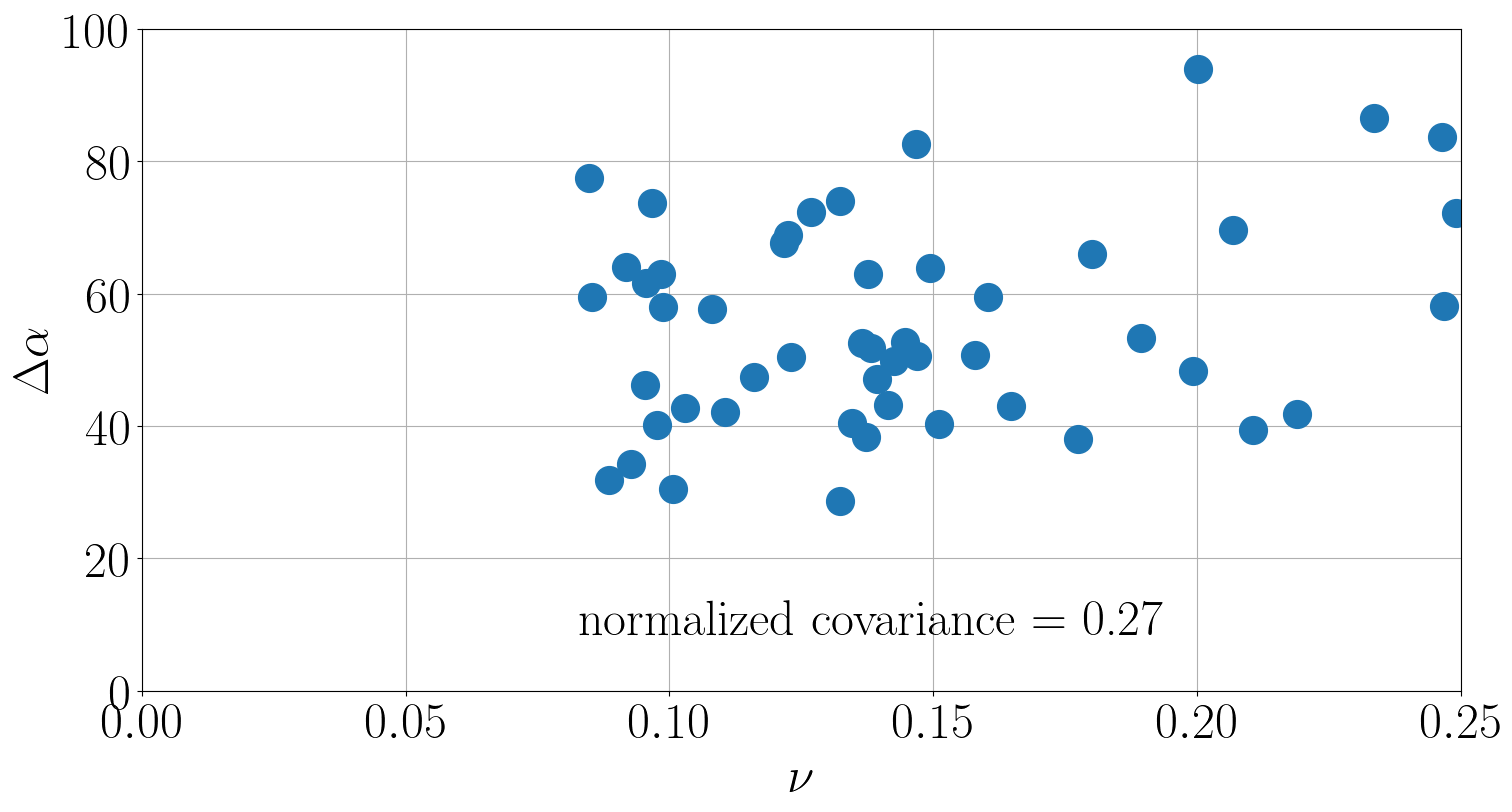}
 \includegraphics[width=0.985\columnwidth]{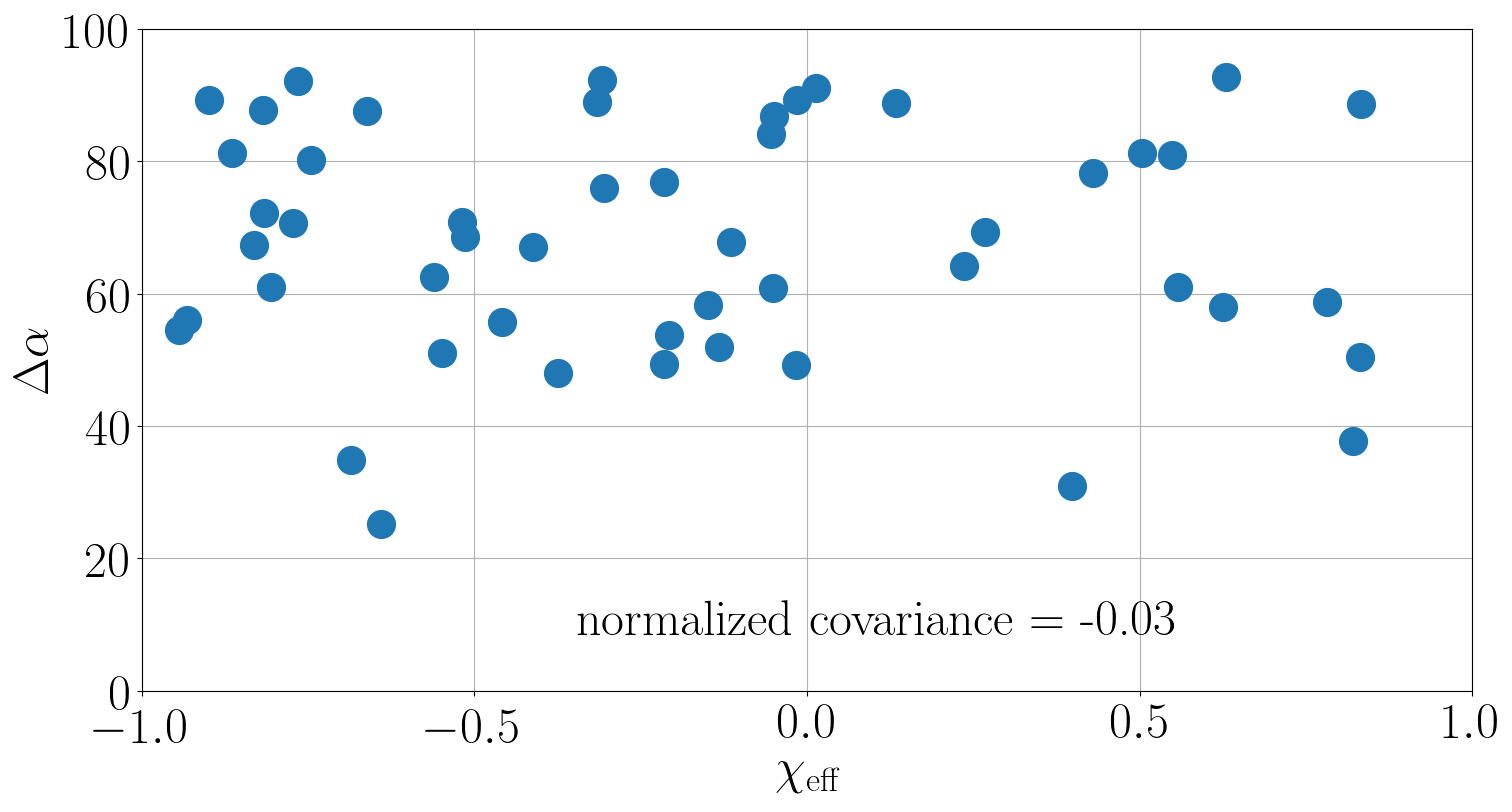}
\caption{90\% confidence interval ($\Delta \alpha$) for $\alpha$ versus (top) $\mathcal{M}_{c}$, (middle) $\eta$, (bottom) $\chi_{\text{eff}}$. The dependencies are quantified by their normalized covariances (top) $(0.52)$, (middle) $(0.27)$, and (bottom) $(-0.03)$. Constraints on $\alpha$ improve with decreasing $\mathcal{M}_{c}$ and $\eta$, while the constraints are only weakly dependent on $\chi_{\text{eff}}$. The result suggests that an astrophysical distribution favoring lower mass black-hole mergers implies a cleaner constraint on the horizon effect by LIGO-Virgo detections.}
\label{fig:vary}
\end{figure}

\noindent
The area theorem discussed in Sec.~\ref{ssec:application} motivate us to search for $\alpha$ at order 1. Therefore, it is sufficient to set the prior of $\alpha$ to be uniformly distributed within $-50 < \alpha < 50$. We investigate the constraints on $\alpha$ of different kinds of binary black holes by Bayesian analysis. Eq.~\eqref{eq:57com} indicates that constraints on $\alpha$ are related to total mass $M$, symmetric mass ratio $\nu$ and spin parameters $\chi_{1}$ and $\chi_{2}$. To investigate the dependencies between the above parameters and constraints on $\alpha$, we fix all inclinations (0\textdegree), sky locations (longitude = 120\textdegree, latitude = -70\textdegree) and distances (400 Mpc) of the simulated gravitational waves events in this section. 
We simulate 3 sets of 50 gravitational waves events: (1) chirp mass $\mathcal{M}_{c} = (m_{1}m_{2})^{3/5}/M^{1/5}$ uniformly distributed from $5 M_{\odot}$ to $35 M_{\odot}$, $\chi_{\text{eff}}= 0.9$ and $\nu = 0.16$, (2) $\nu$ distributed from $0.08$ to $0.25$, $\mathcal{M}_{c} = 30 M_{\odot}$ and $\chi_{\text{eff}}= 0.9$, (3) $\chi_{1}$ and $\chi_{2}$ uniformly distributed from $-0.9$ to $0.9$, $\mathcal{M}_{c} = 30 M_{\odot}$ and $\nu = 0.16$.
For each event, a posterior distribution of $\alpha$ is obtained, and the width ($\Delta \alpha$) of the 90\% confidence interval of $\alpha$ is calculated from the posterior distribution. The results are summarized in Fig.~\ref{fig:vary}, where the variables $\chi_{1}$ and $\chi_{2}$ are collected into a single variable $\chi_{\text{eff}}$. Fig.~\ref{fig:vary} shows that constraints on $\alpha$ improve with decreasing $\mathcal{M}_{c}$ and $\eta$, while the constraints are only weakly dependent on $\chi_{\text{eff}}$. As demonstrated in Fig.~\ref{fig:vary}, the dependencies are quantified by the normalized covariances between the variables and $\Delta \alpha$. Among the three variables, the constraint of $\alpha$ depends most on $\mathcal{M}_{c}$. Therefore, keeping $\eta$ fixed, a cleaner constraint on the horizon effect can be obtained by analyzing a lower mass binary black-hole merger, despite a lower \ac{snr}. The constraint on the horizon effect from multiple LIGO-Virgo detections (see Sec.~\ref{ssec:stack}) depends on the astronomical source mass distribution. For example, an astrophysical distribution favoring lower mass black-hole mergers implies a cleaner constraint on the horizon effect by LIGO-Virgo detections.

\subsection{Stacking posteriors of multiple events} \label{ssec:stack}
\noindent
The constraint on $\alpha$ from a single event may not be strong enough to prove or rule out any theory. Stacking posteriors of different events can give a much better constraint if there is no systematic bias in the posteriors~\citep{huan2016sequential}. Suppose there are $n$ independent detection events $\{H_{1},H_{2},...,H_{n}\}$. For each event $H_{i}$, a posterior distribution is obtained to give a probability density function $P(\alpha|H_{i})$. The multiple events posterior $P(\alpha|H_{1},...,H_{n})$ can be calculated by

\begin{equation} \label{eq:multiple}
P(\alpha|H_{1},...,H_{n}) \propto P(\alpha) \prod_{i=1}^{n} P(H_{i}|\alpha) \propto \prod_{i=1}^{n} P(\alpha|H_{i}).
\end{equation}
Note that we make use of the uniform prior of $\alpha$, $P(\alpha) = \textrm{const}$.

\begin{figure}
 \includegraphics[width=\columnwidth]{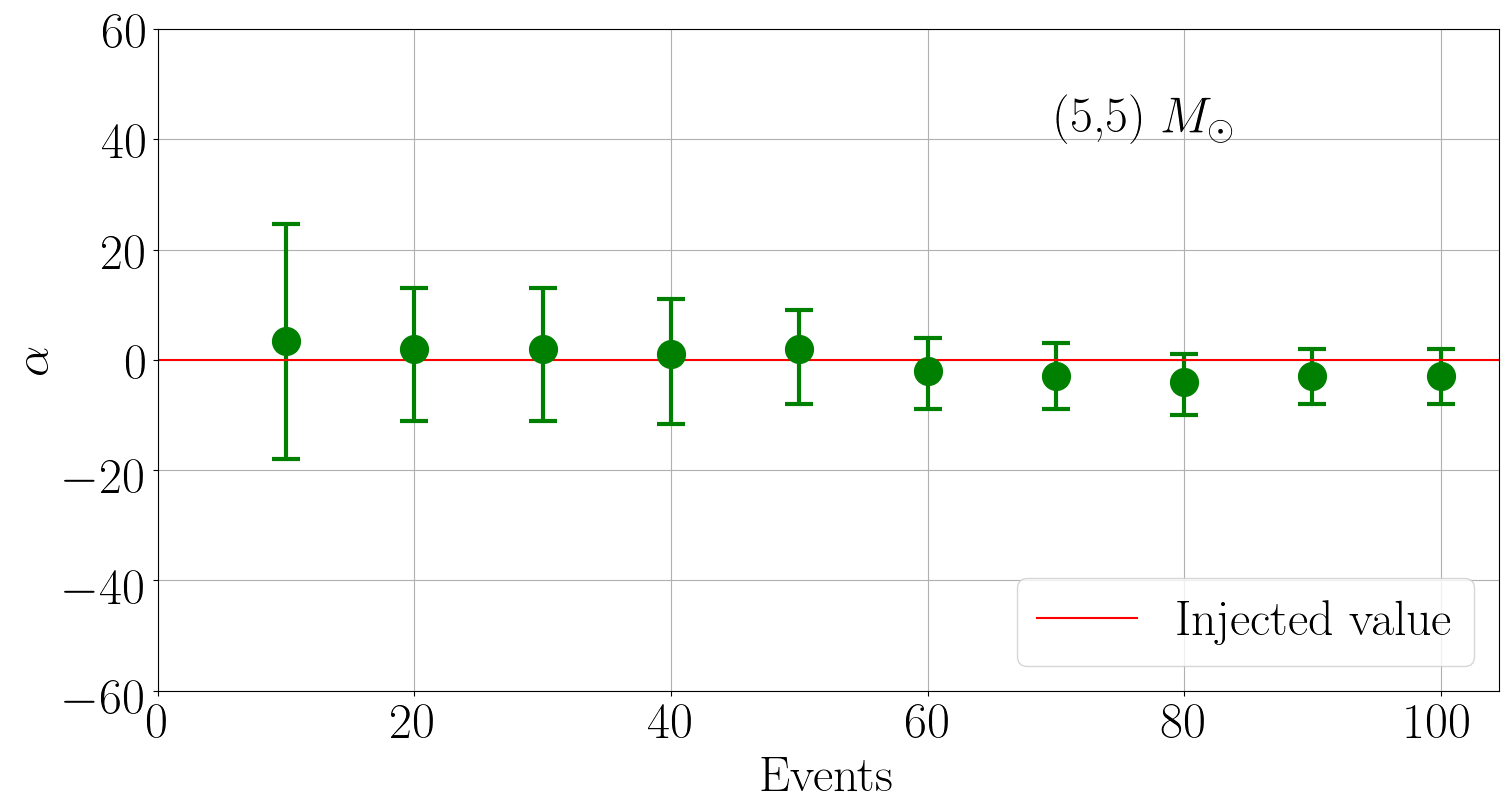}
 \includegraphics[width=\columnwidth]{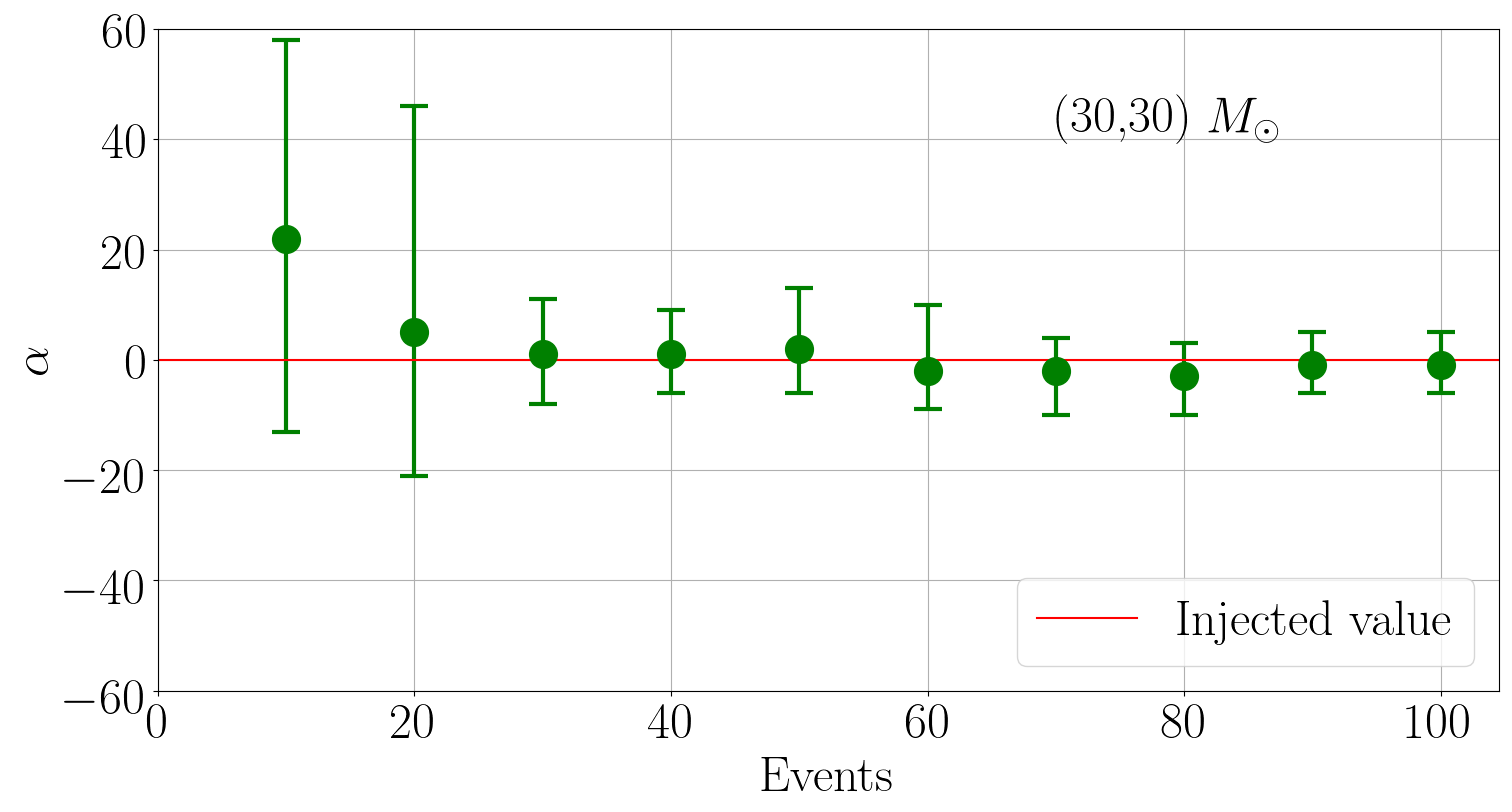}
\caption{Median (dots) and 90\% confidence interval (cross hairs) for $\alpha$ as a function of number of events. Injections were performed using $\alpha=0$ (red line), distributed uniform in $\chi_{1}$ and $\chi_{2}$, isotropic in sky location and uniform in volume, with (top) source masses (5,5) $M_{\odot}$ distributed from 100 Mpc to 200 Mpc, (bottom) source masses (30,30) $M_{\odot}$ distributed from 100 Mpc to 600 Mpc. As the number of events increases, constraints on $\alpha$ improve and approach $\alpha = 0$.}
\label{fig:posterior}
\end{figure}

In Sec.~\ref{ssec:dependency}, we show that the dependency between the chirp mass and the constraint on $\alpha$ is stronger than between the mass ratio and the constraint. To understand the quantitative behavior of the constraint on $\alpha$, we decided to investigate constraints on $\alpha$ from multiple events with different total masses while keeping the mass ratio fixed. We simulate multiple $\alpha=0$ waveforms which are distributed uniform in $\chi_{1}$ and $\chi_{2}$, isotropic in sky location and uniform in volume. Binaries with source masses (5,5) $M_{\odot}$ distributed from 100 Mpc ($\textrm{redshift} \sim 0.02$) to 200 Mpc ($\textrm{redshift} \sim 0.05$) are simulated to investigate an optimistic case since Fig.~\ref{fig:vary} shows that we can constrain $\alpha$ better for smaller chirp mass. Additionally, binaries with source masses (30,30) $M_{\odot}$ from 100 Mpc ($\textrm{redshift} \sim 0.02$) to 600 Mpc ($\textrm{redshift} \sim 0.15$) are simulated to investigate how much can we constrain $\alpha$ in a higher source mass case. The source masses of most of the future \ac{ligo} detections of binary black hole mergers are expected to be in between (5,5) $M_{\odot}$ and (30,30) $M_{\odot}$~\citep{belczynski2016compact}. Priors of $\alpha$ are enlarged to be uniformly distributed in $-500 < \alpha < 500$ to allow larger fractional deviation from general relativity (compared to the prior in Sec.~\ref{ssec:dependency}) during the Bayesian analysis, which helps avoid truncation effects of the posterior when the horizon effect is weak. The waveforms are analyzed using \texttt{LALInference}. Fig.~\ref{fig:posterior} shows the constraints on $\alpha$ of the two sets of simulated events, where the constraints of multiple events are calculated by Eq.~\eqref{eq:multiple}.

In our simulations, 100 gravitational-wave events of (5,5) and (30,30) $M_{\odot}$ binaries constrain $\alpha$ within $-2.8^{+3.3}_{-4.9}$ and $-1.1^{+5.6}_{-4.9}$ (90\% confidence interval) respectively. The better constraint of the (5,5) $M_{\odot}$ (lower chirp mass) case confirms the relation in Sec.~\ref{ssec:dependency}. The width ($\Delta \alpha$) of the 90\% confidence interval of $\alpha$ is $\sim 10$. The independence of the simulated events suggest that the multiple events posteriors can be approximated by a gaussian distribution, which implies $\Delta \alpha \propto 1/\sqrt{N}$, where $N$ is the number of events. When $N \sim 10000$, $\Delta \alpha \sim 1$, suggesting that it is possible to prove the area theorem $\alpha > -1$ (Sec.~\ref{ssec:areatheorem}) at 90\% confidence level. The number of events required to support/disprove a modified theory depends on the corresponding horizon effects parameter in the theory (see Sec.~\ref{ssec:application}). Approximately, 100 detections of gravitational waves can constrain $\gamma$ and $\Xi$ to $\gamma > -0.99$ and $ -150 M_{\odot}^{2} < \Xi < 150 M_{\odot}^{2}$ respectively. The constraint rules out a range of parameters in those theories, which provides a ground for further theoretical research.

\section{Discussion} \label{sec:dis}
\noindent
In this paper, we have developed a practical connection between horizon effects predicted by abstract theories and observable gravitational waves. During the inspiral phase of a binary black hole coalescence, horizon effects can be separated and parameterized by mass-growth parameters $\alpha_{1}, \alpha_{2}$ and spin-growth parameters $\beta_{1}, \beta_{2}$ to describe the mass and spin flux deviations on black hole 1,2 respectively. With the minimal parameterization $\alpha_{1} = \alpha_{2} =  \beta_{1} = \beta_{2} = \alpha$, we show that the 90\% confidence interval of $\alpha$ can be constrained to $\Delta \alpha \sim 10$ with 100 gravitational waves detections from binary black-hole mergers by the Advanced LIGO-Virgo detector network. 
If the modified gravity theory further satisfies the condition where the predicted horizon effect correction dominates over other types of corrections to the geodesic motions, then it is suitable to test the theory by the constraints on the horizon effect. For example, the black hole area theorem, scalar-tensor-vector gravity thermodynamics and quantum corrections to black-hole entropy are suitable to be tested or constrained by their horizon effects.

The parameterized test proposed in this paper can be considered as a subset of \ac{ligo}'s parameterized test~\citep{scientific2016tests}, but aims at separating and measuring the horizon effect alone. Theoretically, \ac{ligo}'s parameterized test is general enough to capture a large set of modified gravity theories. However, the possibility where horizon effect corrections dominate in some theories motivates us to study the slightly more specific parameterized horizon effect instead of considering the full generic test. Although only the parameterization $\alpha_{1} = \alpha_{2} =  \beta_{1} = \beta_{2} = \alpha$ has been considered in our study, a generic $\alpha_{1}, \alpha_{2}, \beta_{1}$ and $\beta_{2}$ as a function of the parameters of the binary black holes  is also possible in some modified gravity theory. The different functional form of $\alpha_{1}, \alpha_{2}, \beta_{1}$ and $\beta_{2}$ can lead to a significantly different constraint. In principle, if a generic parameterized test has been done on a detected signal jointly on the 2.5\ac{pn} and 3.5\ac{pn} phase terms, it is possible to reinterpret the posterior, and search for a specific functional form of degeneracies predicted by the horizon effect, but extra efforts should be spent to avoid bias introduced by the choice of prior probability distributions~\citep{vitale2017impact}. Instead, it is cleaner to constrain horizon effects directly by the method proposed in the paper. 

It is possible that extra corrections to general relativity other than those from the horizon effect correction exist. Depending on the functional form of the extra corrections, we may not be able to discriminate the extra corrections from the horizon effect due to degeneracies between the parameters of the horizon effect and parameters describing these extra effects. Therefore, the horizon effect parameterization is not guaranteed to be a fully model-independent way to extract the information of black-hole horizons. Instead, we suggest that the horizon effect parameterization can be used to test a modified gravity theory only if the theory predicts dominant horizon effect correction over other types of corrections to general relativity. Theories falling into this category can be ruled out by comparing the predicted horizon effects with the constraints obtained by gravitational waves detections.

When extracting the horizon effects from real signals, cautions should be taken to avoid interpreting systematic errors as a violation of general relativity. Since the TaylorF2 model is inaccurate in the late-inspiral-merger-ringdown band, systematic errors arise naturally when a real signal is analyzed by the inaccurate model~\citep{jolien}. Applying a low-pass filter to the remove the late-inspiral-merger-ringdown signal from the data helps to ensure that the violation detected does not come from a misinterpretation of the late-inspiral, merger and ringdown waveform. However, the low-pass filter reduces the strength of the signal, which may weaken the constraint of the horizon effect, and systematic bias can be caused by the abrupt frequency cut~\citep{mandel2014parameter}. On the other hand, improper calibrations of detectors can also induce systematic errors~\citep{abbott2017calibration}. Without proper treatments, the systematic errors can exceed the statistical errors when we try to extract weak horizon effects from real signals~\citep{hughes2018bound}. To establish a rigorous constraint on the horizon effects, studies should be conducted in the future to understand the influence of various systematic errors to the constraint.

Even though the analytical expression of the horizon flux is used to analyze the detectability of the horizon effects in this paper, there are a few theoretical subtleties in the topic. While following similar derivation steps as~\citep{isoyama2017post}, we obtain a slightly different numerical expression of the TaylorF2 horizon effects in Eqs.~\eqref{eq:fluxcoe} and \eqref{eq:57nomodcom} (see Appendix~\ref{modifiedflux} for details). 
Besides, Chatziioannou et al.~\citep{chatziioannou2016improved} point out a disagreement with the extreme mass ratio result in~\citep{tagoshi1997post}. We do not use the unsettled \ac{pn} order expression in our derivation. Moreover, we do not consider the full correction from a self-consistent theory in Sec.~\ref{ssec:application}. Other corrections including metric perturbation corrections and gravitational flux corrections could either amplify or reduce the horizon effect predicted by thermodynamics. Due to the uncertainties, rather than positioning our result as a rigid reference, the paper aims at investigating the possibility of constraining horizon effects using advanced gravitational-wave detectors and the potential of applying a constraint to theories. However, even if the numerical details are not accurate, the effects are expected to be of a similar order of magnitude. 

Assuming that the interpretation in Sec.~\ref{ssec:application} is correct, the area theorem requires approximately $\sim$ 10000 events to give a reasonable constraint at 90\% confidence level depending on the masses of black holes. \ac{ligo} may not be able to detect that large amount of events~\citep{belczynski2016compact}. However, Einstein Telescope is more sensitive than \ac{ligo}, especially at lower frequencies, which means signals can be detected starting from a lower frequency~\citep{sathyaprakash2011scientific,sathyaprakash2012scientific}. Thus, the constraints on $\alpha$ can also be improved by detecting longer signals. Together with the potentially higher event rate~\citep{sathyaprakash2012scientific}, we may have enough events in Einstein Telescope to test area theorem in the foreseeable future. On the other hand, if the strongly-spinning, supermassive black-hole binaries are considered, the LISA detector can pose a reasonable constraint on the horizon effect even by the detection of a single event~\cite{maselli2018probing}. However, the near-horizon correction predicted by a specific modified gravity theory may be mass-dependent, such as the $1/A$ dependence of $\alpha$ of the quantum corrections to black-hole entropy. The theory predicts a stronger correction when the area of the black hole is smaller, which means a lower mass of the black hole. Thus, instead of using the supermassive black-hole binaries detectable by the LISA detector, this kind of theories can be better constrained by the advanced LIGO-Virgo detector network or the Einstein Telescope. 

Despite all the exciting future outlook, the work in this paper has demonstrated that constraints can still be posed on the horizon effects of modified gravity theories using the Advanced LIGO-Virgo detector network. The results motivate us to further investigate the theoretical ground of horizon effects in different modified gravity theories and extend signatures of horizon effect modification to suitable models which can describe late inspiral, merger or ringdown phase accurately~\citep{pan2008data}. Ultimately, constraints on horizon effects can be improved by extracting the effects from full inspiral-merger-ringdown signals.

\section*{Acknowledgement}
The work described in this paper was partially supported by a grant from the Research Grants Council of Hong Kong CUHK 24304317 and by the Direct Grant for Research from the Research Committee of the Chinese University of Hong Kong.

\appendix

\section{Difference between event horizons and apparent horizons during the inspiral phase} \label{sec:aheh}
\noindent
Consider a spherically symmetric black hole which is growing at a constant rate. The difference between the \ac{eh} radius and the \ac{ah} radius in a time slice is given by~\citep{nielsen2010spatial}
\begin{equation}
\delta r = r_{EH} - r_{AH} = 2\frac{\dot{m_{i}}}{\kappa}  + O(\dot{m_{i}}^{2}), 
\end{equation}
where $m_{i},\dot{m_{i}}$ denote the Misner-Sharp mass and its time derivative (with respect to ingoing Eddington-Finkelstein coordinate) of the black hole, and $\kappa$ denotes the dynamical surface gravity~\citep{nielsen2008dynamical}. 

For a typical binary black holes detected by \ac{ligo}, $\delta m_{i}/m_{i} \sim 10^{-6}$~\citep{isoyama2017post}, timescale $T \sim 1s$, $m_{i} \sim M_{\odot}$, $\kappa \sim 1/m_{i}$ implies
\begin{equation}
\frac{\delta r}{r_{AH}} \sim \frac{m_{i}}{T} \frac{\delta m_{i}}{m_{i}} \sim 10^{-12},
\end{equation}
which is negligible unless we can probe the horizon at that scale. Note that the slowly varying approximation $O(\dot{m_{i}}^{2}) \sim 0$ is valid only for the inspiral phase. 

\section{Fully modified horizon flux term} \label{modifiedflux}
\noindent
In this section, we follow the logic of~\citep{isoyama2017post} to compute the full modified horizon flux term.

Define the following symbols for computation convenience,

\begin{equation}
\Delta = -\sqrt{1-4\nu},
\end{equation}

\begin{equation}
S_{l} = \frac{M^{2}}{4}(1+\Delta)^{2} \chi_{1} + \frac{M^{2}}{4} (1-\Delta)^2 \chi_{2},
\end{equation}

\begin{equation}
\Sigma_{l} = -\frac{M^{2}}{2}(1+\Delta) \chi_{1} + \frac{M^{2}}{2} (1-\Delta) \chi_{2}.
\end{equation}

Integration of Eqs. (46) and (47) of~\citep{chatziioannou2016improved} gives individual mass and spin variations $\delta m_{1}$, $\delta m_{2}$, $\delta J_{1}$ and $\delta J_{2}$ as a function of \ac{pn} velocity $x$. With the parameters $\alpha_{1}, \alpha_{2}, \beta_{1}, \beta_{2}$ introduced in Sec.~\ref{ssec:modpara}, the variations are translated into $\delta M, \delta \nu, \delta S_{l}, \delta \Sigma_{l}$ at their leading \ac{pn} order, where $\delta$ means the deviation from the initial value, the \ac{pn} velocity $x = (\pi M f)^{(1/3)}$ is also modified due to the mass modification,

\begin{equation}
\delta M = \frac{1}{56} M [ (1+\alpha_1)  C_{m1}  \sigma_1 + (1+\alpha_2)  C_{m2}  \sigma_2 ]  x^7,
\end{equation}

\begin{equation}
\delta \nu = \frac{1}{56} \nu [ (1+\alpha_1)  C_{\nu 1}  \sigma_1 + (1+\alpha_2)  C_{\nu 2}  \sigma_2 ]  x^7,
\end{equation}

\begin{equation}
\delta S_{l} = \frac{1}{32} M^2 [ (1+\beta_1)  C_{S 1}  \sigma_1 + (1+\beta_2)  C_{S 2}  \sigma_2 ]  x^4,
\end{equation}

\begin{equation}
\delta \Sigma_{l} = \frac{1}{32} M^2 [ (1+\beta_1)  C_{\Sigma 1}  \sigma_1 + (1+\beta_2)  C_{\Sigma 2}  \sigma_2 ]  x^4,
\end{equation}

\begin{equation}
\frac{\delta x}{x} = \frac{1}{3} \frac{\delta M}{M}, 
\end{equation}

\begin{equation}
\begin{aligned}
\sigma_1 ={}& \chi_{1} (1+3\chi_1 ^2), \\
\sigma_2 ={}& \chi_{2} (1+3\chi_2 ^2), \\ 
C_{m1} ={}& -(1+\Delta) \nu + (3+\Delta) \nu^2, \\
C_{m2} ={}& -(1-\Delta) \nu + (3-\Delta) \nu^2, \\
C_{\nu 1} ={}& (1+\Delta) \nu - 2 (2+\Delta) \nu^2, \\
C_{\nu 2} ={}& (1-\Delta) \nu - 2 (2-\Delta) \nu^2, \\
C_{S1} ={}& -(1+\Delta) \nu + (3+\Delta) \nu^2, \\
C_{S2} ={}& -(1-\Delta) \nu + (3-\Delta) \nu^2, \\
C_{\Sigma 1} ={}& (1+\Delta) \nu - 2 \nu^2, \\
C_{\Sigma 2} ={}& -(1-\Delta) \nu + 2 \nu^2. \\
\end{aligned}
\end{equation}
Note that $m_1 < m_2$, $m_1 = M (1 + \Delta)/2$, $m_2 = M (1 - \Delta)/2$.

We denote the energy and the energy flux of the binary without horizon effects by $E$ and $F$ respectively. The first two leading \ac{pn} energy terms $E_{0}, E_{1}$, the leading spin-orbit coupling energy $E^{SO}_{0}$, the first two leading \ac{pn} energy flux terms $F_{0}, F_{1}$ and the leading spin-orbit coupling flux term $F^{SO}_{0}$ are summerized below~\citep{buonanno2009comparison,blanchet2014gravitational}:
\begin{equation} \label{eq:E0}
E_{0} = -\frac{M \nu}{2} x^2,
\end{equation}  
\begin{equation} \label{eq:E1}
E_{1} = -\frac{M \nu}{2} (-\frac{3}{4} - \frac{1}{12} \nu) x^4,
\end{equation}
\begin{equation} \label{eq:ESO}
E^{SO}_{0} = -\frac{\nu}{2 M} (\frac{14}{3} S_{l} + 2\Delta \Sigma_{l}) x^5,
\end{equation}
\begin{equation} \label{eq:F0}
F_{0} = \frac{32}{5} \nu^2 x^{10},
\end{equation}
\begin{equation} \label{eq:F1}
F_{1} = \frac{32}{5} \nu^2 (-\frac{1247}{336} - \frac{35}{12}\nu) x^{12},
\end{equation}
\begin{equation} \label{eq:FSO}
F^{SO}_{0} = \frac{32}{5} \frac{\nu^2}{M^{2}} (-4 S_{l} - \frac{5}{4} \Delta \Sigma_{l}) x^{13}.
\end{equation}
By substituting $M/\nu/S_{l}/\Sigma_{l} \rightarrow M/\nu/S_{l}/\Sigma_{l} + \delta M/\delta \nu/\delta S_{l}/\delta \Sigma_{l}$, the mass and spin variations induce an energy variation $\delta E$ from Eqs.~\eqref{eq:E0},~\eqref{eq:ESO} and an energy flux variation $\delta F$ from~\eqref{eq:F0},~\eqref{eq:FSO} at 3.5\ac{pn} order,
\begin{equation}
\begin{aligned}
\delta E ={}& -\frac{1}{2} (M\delta \nu + \nu \delta M + 2 M \nu \frac{\delta x}{x}) x^2\\
&- \frac{1}{2} \frac{\nu}{M} (\frac{14}{3} \delta S_{l} + 2 \Delta \delta \Sigma_{l}) x^5,
\end{aligned}
\end{equation}
\begin{equation}
\begin{aligned}
\delta F ={}& \frac{32}{5}(2 \nu \delta \nu + 10 \nu^{2} \frac{\delta x}{x} )x^{10}\\
& + \frac{32}{5} \frac{\nu^2}{M^2} (-4 \delta S_l - \frac{5}{4} \Delta \delta \Sigma_{l}) x^{13}.
\end{aligned}
\end{equation}

Since the masses of the black holes are varying, it is natural to add masses $m_{1}$, $m_{2}$ together with $E + \delta E$ to be the total energy of the system $E_{\text{total}}$,
\begin{equation}
E_{\text{total}} = E + \delta E + m_{1} + m_{2}.
\end{equation}
Effectively, the energy balance equation $dE_{\text{total}}/dt = -F-\delta F$ can be written as 
\begin{equation}
\bigg( \frac{\partial (E + \delta E) }{\partial t} \bigg)_{m_{1},m_{2},J_{1},J_{2}}  = -F_{\text{eff}} (x),
\end{equation}
\begin{equation}
F_{\text{eff}} = F + \delta F + (1 + \Gamma_1) F_{H1} + (1 + \Gamma_2) F_{H2},
\end{equation}
\begin{equation}
F_{H1/2} = (1+\alpha_{1/2}) \bigg \langle \frac{dm_{1/2}}{dt} \bigg \rangle,
\end{equation}
where $dm_{1/2}/dt$ can be calculated by Eq. (43) in~\citep{chatziioannou2016improved}. The factors $\Gamma_1, \Gamma_2$ are introduced (to the leading order) to describe the energy change with respect to the mass and spin absorption,
\begin{equation}
\Gamma_{1/2} =  \bigg(\frac{\partial E_{0}}{\partial m_{1/2}} \bigg)_{m_{2/1},\chi_{1},\chi_{2}} + \frac{1}{\Omega m_{1/2}^{2}} \bigg(\frac{\partial E^{SO}_{0}}{\partial \chi_{1/2}} \bigg)_{\chi_{2/1},m_{1},m_{2}},
\end{equation}

\begin{equation} \label{eq:gamma1}
\Gamma_{1} = \left( -\frac{3}{4} + \frac{3 \Delta}{4} + \frac{\nu}{6} \right) x^2,
\end{equation}

\begin{equation} \label{eq:gamma2}
\Gamma_{2} = \left( -\frac{3}{4} - \frac{3 \Delta}{4} + \frac{\nu}{6} \right) x^2.
\end{equation}

The TaylorF2 waveform phase without horizon effects can be described by two master equations,
\begin{equation}
\frac{d\Psi}{df} - 2\pi t = 0,
\end{equation}

\begin{equation} \label{eq:master}
\frac{dt}{df} + \frac{\pi M}{3 x^{2}} \frac{dE/dx}{F} = 0.
\end{equation}
Horizon effects are integrated into the formalism by substituting $E \rightarrow E + \delta E$ and $F \rightarrow F_{\text{eff}}$. $F_{H1/2}$ contributes to the 2.5\ac{pn} phase, while $\delta E, \delta F, \Gamma_{1/2} F_{H1/2}, E_{1} F_{H1/2}, F_{1} F_{H1/2}$ contribute to the 3.5\ac{pn} phase. The TaylorF2 phase of the parameterized horizon effect $\Psi^{F2}_{H}$ is:

\begin{widetext}

\begin{equation} \label{eq:modF2_1}
\begin{aligned}
\Psi^{F2}_{H} ={}& \left(1 + 3 \ln \left( \frac{x}{x_{\text{reg}}} \right) \right) \Psi^{F2}_{H,5}(\alpha_{1},\alpha_{2}) + x^{2} \Psi^{F2}_{H,7} (\alpha_{1},\alpha_{2},\beta_{1},\beta_{2}),
\end{aligned}
\end{equation}
where $x_{\text{reg}}$ can be substitued as the innermost stable orbit $1/\sqrt{6}$.

\begin{equation} \label{eq:57com_1}
\begin{aligned}
\Psi^{F2}_{H,5} ={}& (1+\alpha_{1}) C_{5\alpha 1} + (1+\alpha_{2}) C_{5\alpha 2}, \\
\Psi^{F2}_{H,7} ={}& (1+\alpha_{1}) C_{7\alpha 1} + (1+\alpha_{2}) C_{7\alpha 2} + (1+\beta_{1}) C_{7\beta 1} + (1+\beta_{2}) C_{7\beta 2},
\end{aligned}
\end{equation}

\begin{equation} \label{eq:fluxcoe}
\begin{aligned}
C_{5\alpha 1} ={}& \frac{5}{128 \nu } \chi _1 \left(3 \chi _1^2+1\right) \left(\Delta  \nu -\Delta +3 \nu
   -1\right), \\
C_{5\alpha 2} ={}& -\frac{5}{128 \nu } \chi _2 \left(3 \chi _2^2+1\right) \left(\Delta  \nu -\Delta -3 \nu
   +1\right), \\
C_{7\alpha 1} ={}& \frac{5}{14336 \nu } \chi _1 \big(1740 \chi _1^2 \Delta  \nu ^2+4827 \chi _1^2 \Delta  \nu
   +580 \Delta  \nu ^2+1819 \Delta  \nu -4371 \chi _1^2 \Delta -1667 \Delta
   +828 \chi _1^2 \nu ^2 \\
 &+13569 \chi _1^2 \nu +276 \nu ^2+5153 \nu -4371 \chi
   _1^2-1667 \big), \\
C_{7\alpha 2} ={}& -\frac{5}{14336 \nu } \chi _2 \big(1740 \chi _2^2 \Delta  \nu ^2+4827 \chi _2^2 \Delta  \nu
   +580 \Delta  \nu ^2+1819 \Delta  \nu -4371 \chi _2^2 \Delta -1667 \Delta
   -828 \chi _2^2 \nu ^2 \\
 &-13569 \chi _2^2 \nu -276 \nu ^2-5153 \nu +4371 \chi
   _2^2+1667 \big),\\
C_{7\beta 1} ={}& -\frac{15}{4096} \chi _1 \left(3 \chi _1^2+1\right) \left(18 \Delta  \nu -59 \Delta +136
   \nu -59\right), \\
C_{7\beta 2} ={}& \frac{15}{4096} \chi _2 \left(3 \chi _2^2+1\right) \left(18 \Delta  \nu -59 \Delta -136
   \nu +59\right).
\end{aligned} 
\end{equation}

Note that if no modification is made towards general relativity, $\alpha_{1} = \alpha_{2} = \beta_{1} = \beta_{2} = 0$. Eq.~\eqref{eq:57com} reduces to 

\begin{equation} \label{eq:57nomodcom}
\begin{aligned}
\Psi^{F2}_{H,5} ={}&  C_{5\alpha 1} + C_{5\alpha 2}, \\
\Psi^{F2}_{H,7} ={}& C_{7\alpha 1} + C_{7\alpha 2} + C_{7\beta 1} + C_{7\beta 2}.
\end{aligned}
\end{equation}

\end{widetext}

Substituting $\alpha_{1} = \alpha_{2} = \beta_{1} = \beta_{2} = 0$, we successfully reproduce the result of~\citep{isoyama2017post} until Eq.~\eqref{eq:master}, $\Psi^{F2}_{H,5}$ in Eq.~\eqref{eq:57nomodcom} is also the same as that in~\citep{isoyama2017post}. However, the numerical value of $\Psi^{F2}_{H,7}$ in Eq.~\eqref{eq:57nomodcom} is slightly different from that in~\citep{isoyama2017post}. Comparing to~\citep{isoyama2017post}, $\Psi^{F2}_{H,7}$ in Eq.~\eqref{eq:57nomodcom} contains an extra term 
\begin{equation}
\begin{aligned}
& -\frac{5}{14366} [(-1 + \Delta(\nu - 1) + 3\nu)\chi_1 (1 + 3\chi_1^2) \\
& - (1 + \Delta(\nu - 1) - 3\nu )\chi_2 (1 + 3\chi_2^2)], 
\end{aligned}
\end{equation}
where we notice that this extra term can be written as 
\begin{equation}
-\left( \frac{5}{256 x^7 \nu} \right) \left( \frac{\delta M}{M} \right).
\end{equation}

We fail to resolve the origin of this difference. Fortunately, for the parameter range considered in this paper, the difference is negligible. For example, if we substitute $\nu = 0.25$, $\chi_1 = \chi_2 = 0.5$, the extra term is only 0.02\% of the numerical value of $\Psi^{F2}_{H,7}$. If we substitute $\nu = 0.1$, $\chi_1 = \chi_2 = 0.5$, the extra term reduces to 0.006\% of the numerical value of $\Psi^{F2}_{H,7}$. The relatively small value of the extra term assures that it has negligible influence on both the qualitative and quantitative behavior of the constraints analyzed in this paper.

\section{Examples of horizon effect parameterization} \label{sec:paraexamples}
\noindent
In the following, we demonstrate the relation between some specific theories and the horizon effect parameterization.
 
\subsection{Area theorem} \label{ssec:areatheorem}
\noindent
It is well known that if the null energy condition is satisfied, then the areas of black holes are non-decreasing~\citep{chrusiel2001regularity}. During the inspiral phase, the positivity of area growth is shown by Eq. \eqref{eq:area} where $dA/dx$ is always positive. 
By adopting $\alpha_{1} = \alpha_{2} =  \beta_{1} = \beta_{2} = \alpha$, and assuming that the first law of black hole thermodynamics is unmodified, the change of a black hole area can be interpreted as 
\begin{equation} \label{eq:areamod}
\left \langle \frac{dA}{dx} \right \rangle \rightarrow (1 + \alpha) \left \langle \frac{dA}{dx} \right \rangle.
\end{equation}
If $\alpha < -1$ is observed in future gravitational-wave events, the area theorem will be proven wrong. In Sec.~\ref{ssec:stack}, we show that the stacking of gravitational wave constraints of $\alpha$ should be able to tell whether $\alpha < -1$ from observations.

\subsection{Scalar-Tensor-Vector Gravity} \label{ssec:STVG}
\noindent
Scalar-Tensor-Vector gravity introduces an extra scalar field and a vector field with the standard Einstein-Hilbert action, which predicts a corrected temperature and an corrected entropy for a static black hole~\citep{mureika2016black}.
\begin{equation}
\begin{aligned}
T(\gamma) = \frac{1}{2\pi G_{N} m_{i}} \frac{1}{(1+\sqrt{1+\gamma})(1+\gamma+\sqrt{\gamma})},
\end{aligned}
\end{equation}

\begin{equation}
\begin{aligned}
S_{A}(\gamma) ={}& \pi G_{N} m_{i}^{2} (1+\sqrt{1+\gamma})^{2} \\
& - \frac{1}{2} \ln \left( \frac{1}{4\pi G_{N}} \right) \frac{1}{(1 + \gamma + \sqrt{1+\gamma})^{2}},
\end{aligned}
\end{equation}
where $G_{N}$ is the Newtonian gravitational constant, and $\gamma$ is a non-negative real number representing the modification. Note that we use $\gamma$ to denote the $\alpha$ in~\citep{mureika2016black}. $\gamma = 0$ corresponds to unmodified general relativity. 

This correction is calculated for a non-rotating black hole, so it is just an approximate correction towards the more realistic rotating solution. 

It is reasonable to assume black holes still follow the first law of black hole thermodynamics with an modified entropy,

\begin{equation}
T \langle \dot{S}_{A} \rangle = \langle \dot{m_{i}} \rangle - \Omega_{H} \langle \dot{J} \rangle,
\end{equation}
where $T = \kappa/2\pi$, $S_{A} = A/4$ recovers general relativity (see Eq. \eqref{eq:firstlaw}).

For simplicity, assuming that only corrections on the first law contribute to the mass and spin flux corrections, which is not precise in a full modified gravity theory since the perturbation does not follow Teukolsky's equation. The flux corrections can be translated to $\alpha = \alpha_{1} = \alpha_{2} =  \beta_{1} = \beta_{2}$, 
\begin{equation}
\alpha = \frac{T(\gamma) dS_{A}(\gamma)} {T(0) dS_{A}(0)} - 1 = - \frac{\gamma}{1+\gamma+\sqrt{1+\gamma}}.
\end{equation}

$\gamma < 0$ is not considered to be a physical case in~\citep{mureika2016black} since it refers to a complex gravitational charge. However, $\gamma > -1$ is well defined for both the temperature and the entropy, and only observations can tell whether it is physical or not.

Also, note that Mureika et al. derive a zeroth \ac{pn} order correction in the theory~\citep{mureika2016black}, while Moffat suggests that the correction is just phenomenological and it is not necessarily true within the theory~\citep{moffat2006scalar}.

\subsection{Quantum corrections to black-hole entropy} \label{ssec:QG}
\noindent
Even though many quantum gravity theories predict general relativity corrections at the Planck scale, it is possible to observe some corrections at the black hole thermodynamics level. Effective field theory of quantum gravity predicts a logarithmic correction towards the Schwarzschild black-hole entropy~\citep{Carlip20004175,el2016quantum}. Again, the correction is just an approximate correction to the Kerr black hole solution.

\begin{equation}
S_{bh} = S_{BH} + 64\pi^{2} \Xi \ln \left( \frac{A}{A_{QG}} \right),
\end{equation}
where $S_{bh}$ is the corrected entropy, $S_{BH} = A/4$ is the ordinary black-hole entropy, $A_{QG}$ is a quantum gravity area scale, and $\Xi$ represents a massless-particles contribution in the model (refer to~\citep{el2016quantum} for the details). 

Similar to Sec.~\ref{ssec:STVG}, the correction can be translated to $\alpha$.
\begin{equation}
\alpha = \frac{dS_{bh}}{dS_{BH}} - 1 = \frac{256\pi^{2} \Xi}{A}.
\end{equation} 
Note that $\alpha$ is mass dependent, because $1/A = 1/16\pi m_{i}^{2}$.

\bibliographystyle{unsrt}
\bibliography{paper}

\end{document}